# Lossless Compression of Color images using Imperialist competitive algorithm

Abbas Mirzaei SOMARIN [1,*], Mohammad Reza Deldadeh SHIRIN [2]

[1]Department of Computer Engineering, Ardabil Branch, Islamic Azad University, Ardabil, Iran
[2]Department of Computer Engineering, Ardabil Science and Research Branch, Islamic Azad University, Ardabil, Iran

## Abstract

One of the most important reasons of the existence of different types of files with media (audio or video) content, is achieving compression and less size, while preserving quality. In terms of fast transportation of files between equipment and networks and decrease of the required storage space, compression have always been under attention and action. Considering what was mentioned, in general, the concept of compression can be divided into two classes of lossy and lossless. In the lossy method, a part of data is omitted, but in the lossless methods, no data is omitted for compression. At the end of this article, a lossless compression method is presented using Imperialist competitive algorithm for compression. The proposed algorithm tries to achieve a more optimized color for the image color-map, so that it increases the compression rate. The simulation results indicate that the proposed algorithm can perform the compression by 43 percent and show its superiority compared to other similar methods.

## 1-Introduction

Nowadays, with the development of technology, fast and feasible access to a huge mass of information is an undeniable and necessary matter. Considering the large mass of information which is mostly consisted of images, the existence of systems in order to compress and recover images seems necessary (MING-NI WU et al,2006 and S. Battiato et al, 2004). Image compression methods can be divided into two classes of lossy and lossless. Lossless compression algorithms usually implement statistical frequency in a way to show the senders information more briefly and without any errors. Another type of image compression is the lossy or perceptual coding. In lossless image compression, an image file is compressed without any losses in its data. Lossless image compression compresses file sizes in a way that allows more image data to be recovered bit by bit so that the original file can be recovered. The lossless compression methods simply decrease data redundancy. While in lossy compression techniques, practically a part of the original image data is thrown away.

One of the issues in the field of lossless color images compression is the issue of changing the color rows which solving will result in an enormous impact on decreasing image size after compression. An image which has a color-map is also called a thumbnail. Thumbnail uses direct mapping of pixels resolution for color mapping. The color of each pixel is determined using the corresponding amount of the accurate matrix of X as the color mapping indicator.

Hence, it is called a thumbnail. Thus, considering these concepts, with M different colors, M maps with different colors can be obtained. The implication of compression algorithm relies much on the allocation method of the images indicators to the colors of the map (MING-NI WU et al,2006 and S. Battiato et al, 2004). So, by changing the order of different colors in the color-map, the order of the image indicator also changes. This factor has a direct impact on compressing thumbnails, because the more pixels are of lower diversity, the entropy of the image will be decreased and this leads to increasing the image compression rate. Finding the best image indicator order which leads to the best compression rate is an NP-Complete issue (Memon ND and Venkateswaran A, 1996).Thus, in this article, a method based on the imperialist competitive algorithm is proposed for solving the optimization issue of color-map row order and compared to its former similar methods, be capable of increasing the compression rate without loss of the image.

In part 2 of this article, literature and related works are described. In the third part, thumbnails and the issue of color-map row order are investigated. In part 4, the suggested method is proposed. In part 5, the results of simulation and implementation of the suggested method are studied and analyzed.

## 2- Literature

Considering that compression is significantly important without images, this subject has always been under the attention of researchers. Spira et al. (2010) (MING-NI WU et al,2006) proposed a method based on distances between the pixels colors. Their assumption is that an image is built of pixel bodies with similar colors, then they proposed a method using determining the furthest pixel color algorithm for lossless compression. The time complexity of the mentioned article is o(n3). In Niraimathia et al. (2012) (S. Battiato et al, 2004) a method was presented based on particle swarm algorithm for the optimization issue of color map row order. In this article also the issue of color map row order change is first introduced and then it is proved that it is NP-Complete. And then using the particle swarm algorithm, a method is proposed for lossless compression of color images, which tries to create an equilibrium between the compression rate and compression time. In Sebastian et al. (2004) (A.spira and D.Malah, 2001), the optimization issue of color-map row order in thumbnails is discussed and a method is introduced according to the Graph theory to solve this issue. The introduced method tries to find an optimized order in the color-map through the heaviest Hamiltonian route in a weighted graph. Ming et al (2007) (Pinho AJ and Neves AJR, 2004) has introduced a lossless method for compression of color thumbnails using genetic algorithm. In this method, the face image is coded in a permutation manner and the genetic algorithm functions according to the permutation coding. Through simple genetic operators, this method were able to increase the compression rate up to 40 percent. Although compression time remained unchanged. In a research done by Akbari Moghadam et al. (2014) (Memon ND and Venkateswaran A, 1996) the issue of color map order change is discussed and it is proved that it is NP-Complete and then a method were presented for solving the problem above according to divide and conquer algorithm. In this method as well, compression were improved up to 30 percent. In Azizpour et al. (2015) (Battiato S et al, 2007) a method were proposed according to genetic algorithm for lossless compression of color images. In this article, two new methods were proposed for combination activators and changing suitable with color map of the images, which can suggest a suitable order of color map for compression. Sodhakar et al. (2011) (Kuroki N et al, 2004) have proposed a method according to the graph theory in order to compress color thumbnails. In this method, color map of the image indicators are turned into weighted graphs. Then the weighted graph is adjusted to the travelling salesman problem and after that, a suitable order for color maps is obtained for compression. Battiato et al (2005) (Koc B and Arnavut Z, 2011) have introduced a greedy method according to the graph theory. In this method, after the graph of color map is obtained, using a greedy method, the optimum order of color map is sought with the maximum compression rate. Rando et al (2013) (Koc B and Arnavut Z, 2012) have used neural network for finding the suitable order of color map with maximum compression rate. In this method, the problem of changing the order of color map is formulized in a way that the neural networks can be used for solving it. Moreover, in Hog et al. (2010) (JPEG, 2009) a new lossless compression method is presented for images with combining the Particle swarm algorithm and the travelling salesman problem, which they used the graph theory for formulization of the problem and then, they solved the intended problem with the proposed method.

## 3- Statement of the problem

Take image I with mn pixels, in a way that its amount in the ith row and jth column is I (i, j) ∈ {1,2 ,…,M} I (i, j). An m lined table with three columns indicating the corresponding color of each row is taken as P, which indicates the color of each pixel of the image according to the number amount of its row. Consequently, an ordered set of colors is obtained as P= (p1, p2,…,pM) which is the color map of the Mth image (JPEG, 2009). Usage of the optimization algorithm depends highly on the method of image indicator allocation to the colors of the map. With M different colors, different maps are obtained. Finding the optimum color map, a color map which has the lowest image storage size after compression needs to be sought in this vast space. This problem is a NP-Complete problem. Figure 1 shows an image with its pixels. Each pixel is formed with its corresponding RGB colors. Figure 2a shows the color of the image in Figure 1. This table indicates that the image in Figure 1 is only formed of 4 different colors and these pixels only keep the indicator amount of the color map instead of keeping their original color amounts. Thus, the image is stored with color map is it is shown in Figure 2b. Also, if the order of color map were as Figure 5a, the image would be stored like Figure 3b with color map.

| | | | |
|---|---|---|---|
| (100,20,50) | (60,150,200) | (60,150,200) | (140,140,120) |
| (100,20,50) | (30,70,80) | (30,70,80) | (60,150,200) |
| (140,140,120) | (100,20,50) | (60,150,200) | (100,20,50) |
| (100,20,50) | (60,150,200) | (140,140,120) | (100,20,50) |

**Figure 1:** An image with its pixels amounts (MING-NI WU et al , 2006)

0: (100,20,50)
1: (60,150,200)
2: (140,140,120)
3: (30,70,80)

| 0 | 1 | 1 | 2 |
|---|---|---|---|
| 1 | 3 | 3 | 1 |
| 2 | 0 | 1 | 0 |
| 0 | 1 | 2 | 0 |

**a**　　　　**b**

**Figure 2:** An order of color map for the image of Figure 3 (MING-NI WU et al, 2006)

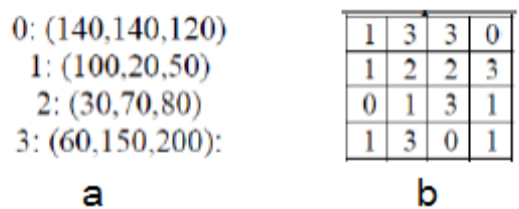

**Figure 3:** An order of color map for the image of Figure 3 (MING-NI WU et al, 2006)

Thus, with different color orders in color maps, the image indicator order also changes. This factor has a direct effect on compressing the thumbnails compression. Because, as close pixels have less differences with each other, the image entropy decreases and leads to decrease of compression rate of image. Finding the best order of image indicators which leads to the best compression rate is an optimization and NP-Complete problem.

**4- Suggested method**

In this part, a new method is presented for compression of color images using imperialist competitive algorithm and then it is explained in detail. The imperialist competitive algorithm like other methods of evolutionary optimization methods starts with the primary population. In this algorithm, each population element is called a country. Countries are divided into two groups of colony and imperialist. Each imperialist, depending on its power, colonizes some imperialist countries and controls them. The imperialistic attraction and competition policy forms the main core of this algorithm.

**4-1 definition of countries (generating the primary answers)**

In order to begin with the proposed algorithm, answers (countries) should be generated randomly and admissibly. This article aims to find color maps that generate minimum size for images after compression comparing to the primary color map. Thus, each answer is different orders of indicators 1 to M, in a way that each one of its numbers points to its corresponding row of the primary color map which its columns indicate the RGB color space indicators (JPEG, 2009). Since each color map includes an order of the image colors and different color maps consider different orders for colors, the best coding method of color maps is the permutation coding method. Assume that the intended image is consisted of 8 different colors. With such an assumption, if we show the colors with $C_i$, coding will have three different indicators of R, G, and B. in fact, figure 4 shows two samples of the countries of the imperialist competitive algorithm.

| CH1 | C1 | C4 | C8 | C2 | C7 | C3 | C5 | C6 |
|-----|----|----|----|----|----|----|----|----|
| CH2 | C6 | C1 | C7 | C3 | C5 | C8 | C2 | C4 |

**Figure 4:** Two samples of coded color maps
(acceptable answers for the imperialist competitive algorithm)

**4-2 assessing the countries suitability**

In this article, in order to assess the suitability of each country, we use entropy. The lesser the entropy, the more power the country has. In order to calculate the quality of each color map, first the amounts of image pixels that are stored in a two dimensional manner are turned into a $1 \times n$. For this, we store the image pixels in a zigzag manner. Then we calculate the pixels differences by order (instead of storing the main amount of pixels, the difference amounts which need much less space are stored). Thus, the difference of the neighboring pixels is obtained through Equation 1, where n is the image size, $S_i$ is the ith pixel of the image which is scanned by a linear scan. Thus, the difference matrix entropy which is also called the first grade entropy is used as the standard function ($fi = Ds$) of the imperialist competitive algorithm.

$$D_s = \sum_{i=1}^{mn-1} |s_i - s_{i-1}| \qquad (1)$$

**4-2 formation of imperials**

In order to begin with the algorithm, there should be some primary imperials. Thus, we should generate Ncountry primary countries in a permutation and random manner. Then Nimp of the best members of this population (countries with the least amount of better suitability) as imperialist. The rest of Ncol countries form colonies which each belongs to one of the imperials. In order to divide the first colonies among imperialists, each imperialist is given a number of colonies suitable with its power. For this, having the cost of all imperialists, their normalization cost is as Equation 2.

$$C_s = \max\{c_i\} - c_n \qquad (2)$$

Where cn is the nth imperialist cost ($D=c$), $\max\{c\ i\}$ is the maximum cost among imperialists, and Cn is the normalization cost of this imperialist. Each imperialist which has the most cost (weaker imperialist in terms of suitability), would have less normalization cost. Having the normalization cost, the relative normalization power of each imperialist is calculated as Equation 3 and according to that, the colonies are divided among imperialists.

$$p_n = \left| \frac{C_n}{\sum_{i=1}^{N_{imp}} C_i} \right| \qquad (3)$$

**4-3 attraction and assimilation policy**

In order to define the attraction policy, we use the similar operator with combination operator in the genetic algorithm. In order to act on the attraction policy, each country is mixed with its imperialist and form a new country. The new country replaces its former country. Countries are chosen according to their attraction probability in order to implement the combination operator upon them (combination with imperialist). Here, we use the single point method for the combination operator. Here, according to the single point method, some colonies are chosen according to their combination (attraction) probability for combination with the imperialist. The combination act is conducted in a way that one point is determined randomly in the imperialist, then all the former elements are exactly copied in the colony from the crop point and a new colony is formed which is replaced with the former one. Thus, the newly generated colony would have similar elements with the imperialist elements. Assuming that we have 8 colonies (the number of image pixels), figure 5 shows an instance of how the combination act is conducted for acting on the attraction policy.

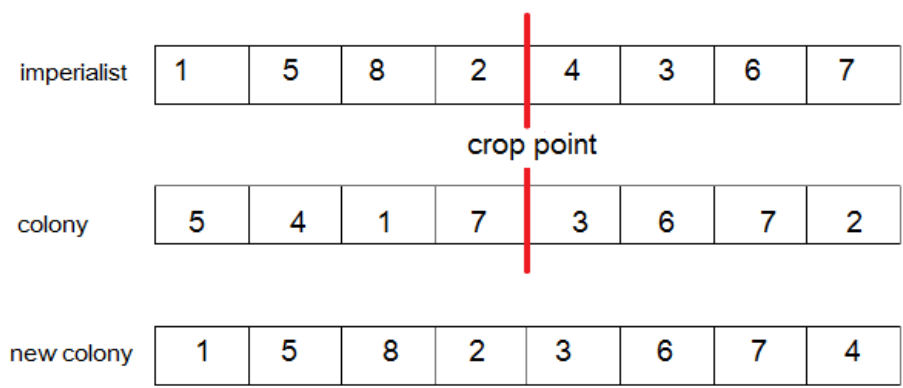

**Figure 5:** An instance of how the combination act is conducted to act upon the attraction Policy

### 4-4 calculating the total power of an imperialist

After the attraction policy was conducted, the imperials' power is calculated. The power of an imperial equals the power of imperialists in addition to a percentage of the total power of its colonies. Thus, according to the suitability amount of each country, the power of each imperial is calculated as equation (4).

$$\text{T.C}_n = f\left(\text{imperialist}_n\right) + \alpha\left(f\left(\text{colonies of empire}_n\right)\right) \qquad (4)$$

Where $\text{T.C}_n$ is the total cost of the nth imperial and $\alpha$ is a positive number which is usually considered between zero and 1, approximate to zero. $\alpha$ being considered small results into the relative equality of the total cost of an imperial with the cost of its central government (the imperialist country), and by increasing $\alpha$, the cost level of colonies of an imperial to be increased in determination of its total cost.

### 4-5 replacing the position of colony with imperialist

After each time the attraction act is conducted, the suitability amount of the countries of each imperial is calculated. If the suitability of the colony be higher than the suitability of the imperialist in the same imperial, the place of colony and imperialist will change in that imperial.

**4-6 imperialist competition and downfall of weak imperials**

In modeling the imperialist competition, we assume that each imperial which cannot add to its power and lose its competition power are the imperials being deleted and the weakest existing imperials. Hereby, in iteration of the algorithm, one or some of the weakest colonies of the imperial are taken and then we start a competition among all imperials for conquering these colonies. The mentioned colonies are not necessarily conquered by the strongest colony, but they are more likely to take control of them. In the imperialist competitive algorithm, an imperial is perceived as deleted when that imperial has lost its colonies.

**4-7 final condition**

The imperialist competitive algorithm continues until an acceptable answer for the compressed image or until the end of all iterations. After a while, all imperials are fallen and we would only have one imperial and other countries will be under the control of this one imperial. The final condition for the imperialist competitive algorithm in this article is considered as bellow:
In n sequential iteration of the algorithm, no refinement were done in the countries suitability. The algorithm iteration reaches the number of the completion of all iterations.

**5-Simulation and result evaluation**

  The suggested algorithm is implemented in this article using MATLAP. Simulations are conducted on 6 sample images, which one time it was conducted with 64 different colors and one time with 16 different colors in the size of 255*255. The sample images are indicated in figure 6.

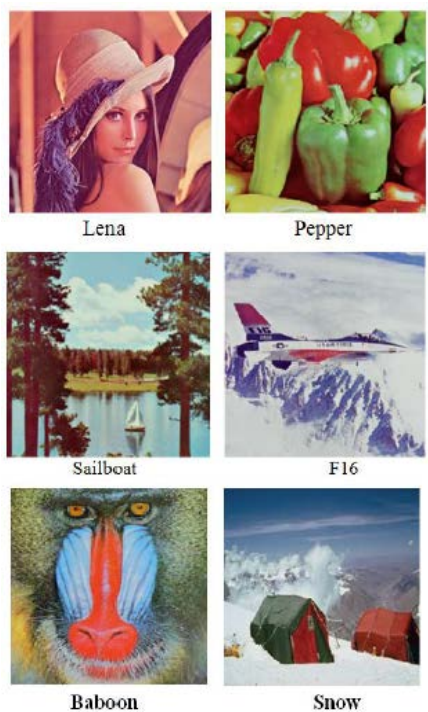

**Figure 6:** Sample images

**5-1 results evaluation according to the compression rate**

The shown chars in figures 7 o 12 indicate the images compression rate when the images are of 64 different colors. The compression rate of the suggested method is compared with a method based on the particle swarm algorithm (P. Niraimathia et al,2012).Image (4-2) shows the compression rate obtained from algorithms on the image Lena, also, image (4-3) shows the compression rate obtained from algorithms on the image Pepper, image (4-4) shows the compression rate obtained from algorithms on the image Sailboat, image (4-5) shows the compression rate obtained from algorithms on the image F16, image (4-6) as well shows the compression rate obtained from algorithms on the image Baboon, and image (4-7) shows the compression rate obtained from algorithms on the image Snow. According to the obtained results in the images above, the compression rate of the suggested algorithm is higher than the particle swarm based method and the suggested algorithm were able to decrease the size of the image with higher compression rate. In average, in all of the 6 images with 64 different colors, the suggested algorithm were able to conduct the compression act with 43 percent compression rate.

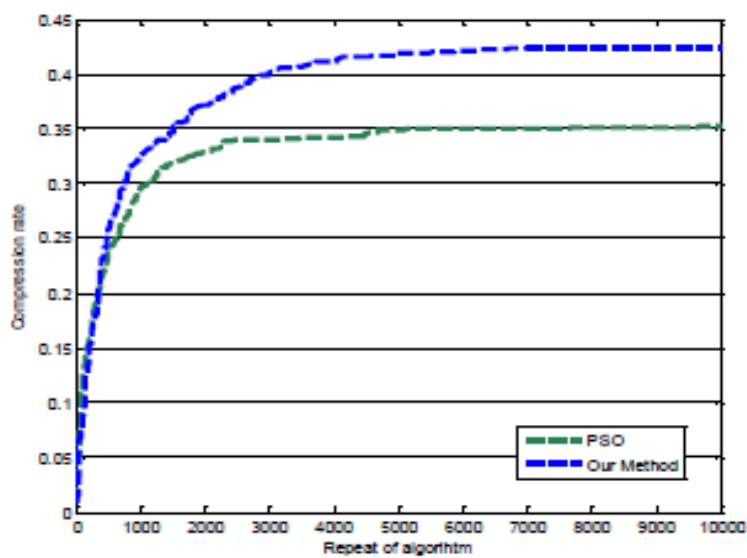

**Figure 7:** Results obtained from the compression rate upon the image Lena (with 64 colors)

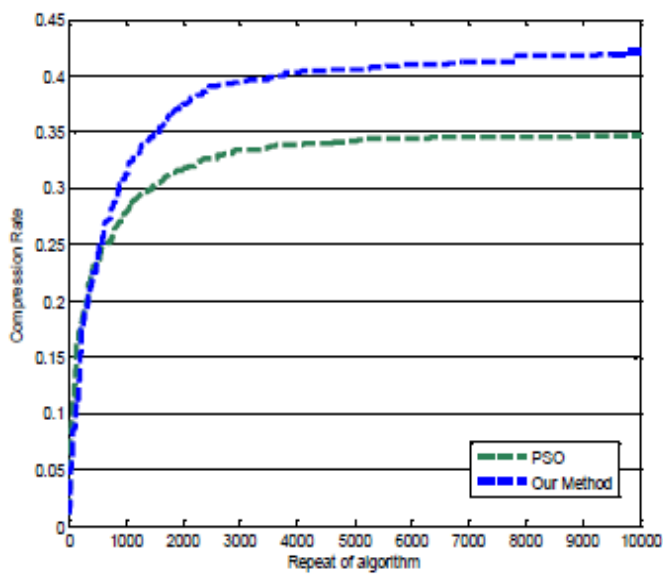

**Figure 8:** Results obtained from the compression rate upon the image Pepper
(with 64 colors)

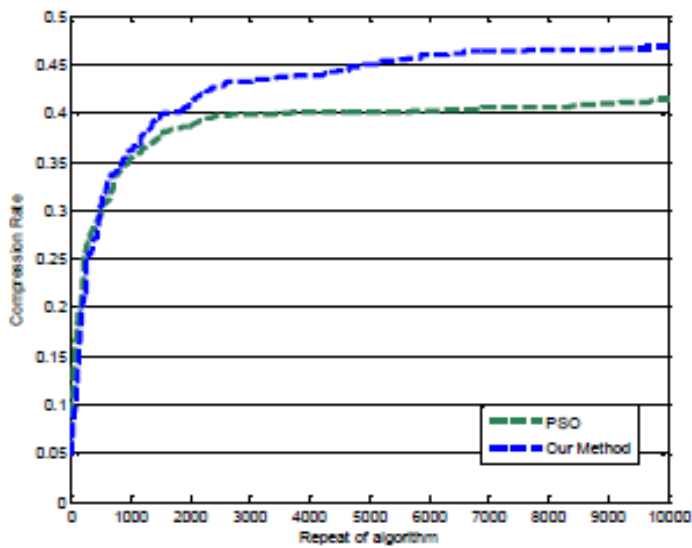

**Figure 9:** Results obtained from the compression rate upon the image Sailboat
(with 64 colors)

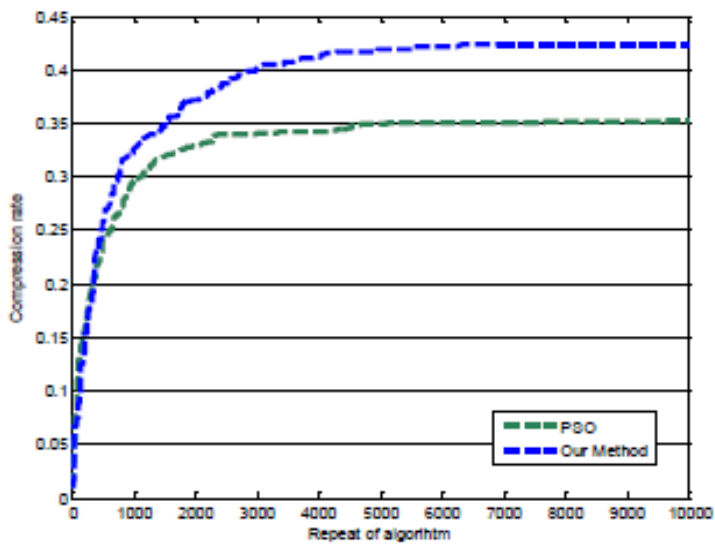

**Figure 10:** Results obtained from the compression rate upon the image F16
(with 64 colors)

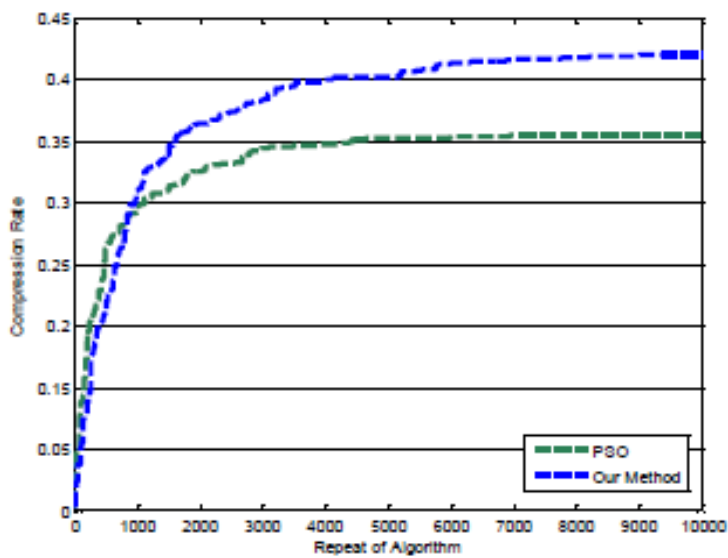

**Figure 11:** Results obtained from the compression rate upon the image Baboon
(with 64 colors)

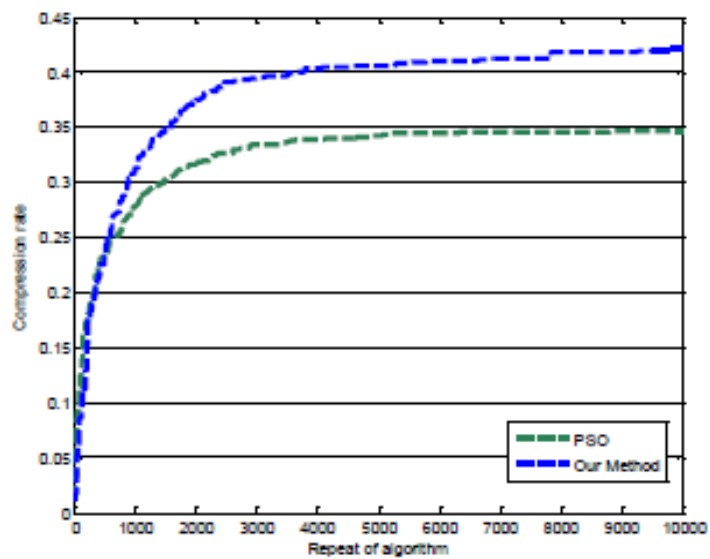

**Figure 12:** Results obtained from the compression rate upon the image Snow
(with 64 colors)

In above figures, the compression rate were conducted on images when they were of 64 colors. Now we conduct the suggested algorithm on the same images, with the difference that these images are of 16 different colors. The shown charts shown in figures 13 to 18 indicate the images compression rate when they have 16 colors. Thus, when the images have 16 different images also the compression rate of the suggested algorithm is better and has increased the compression rate compared to the method based upon particle swarm. In average, in each 6 images with 16 different images, the suggested algorithm were able to do the compression act with 37 percent compression rate.

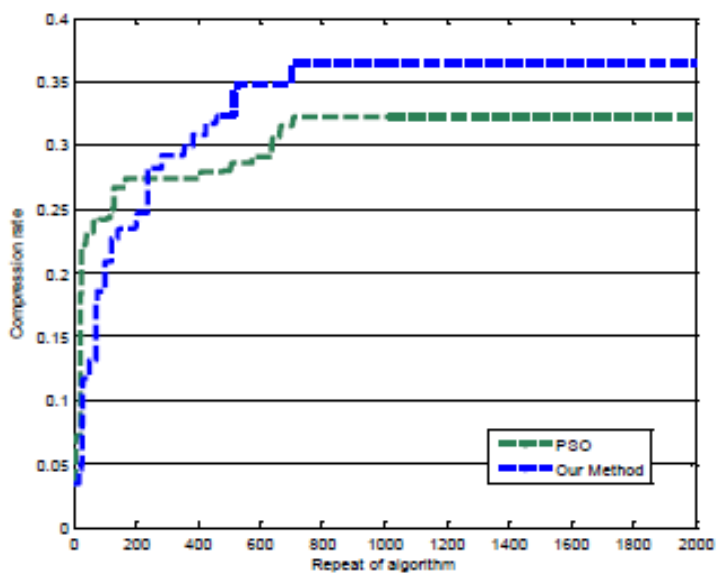

**Figure 13:** Results obtained from the compression rate upon the image Lena
(with 16 colors)

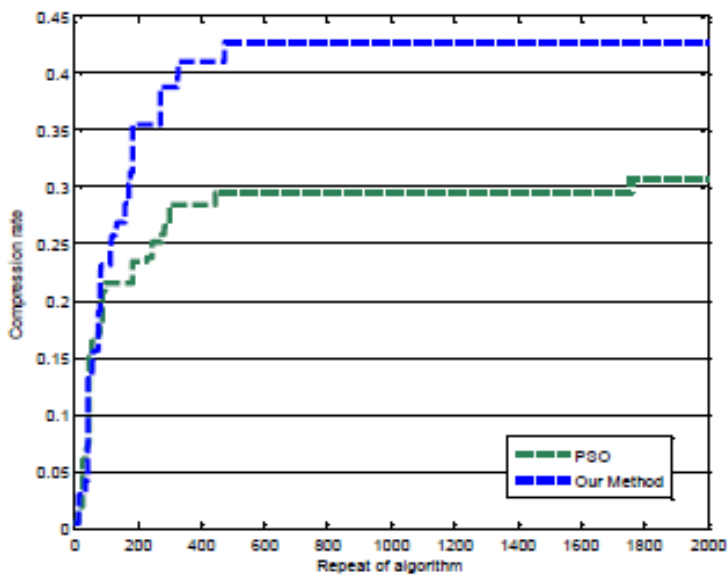

**Figure 14:** Results obtained from the compression rate upon the image Pepper
(with 16 colors)

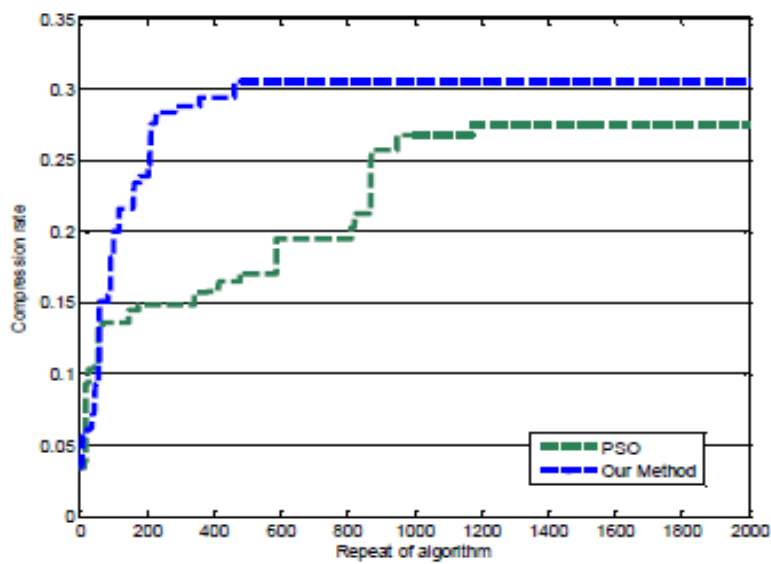

**Figure 15:** Results obtained from the compression rate upon the image Sailboat
(with 16 colors)

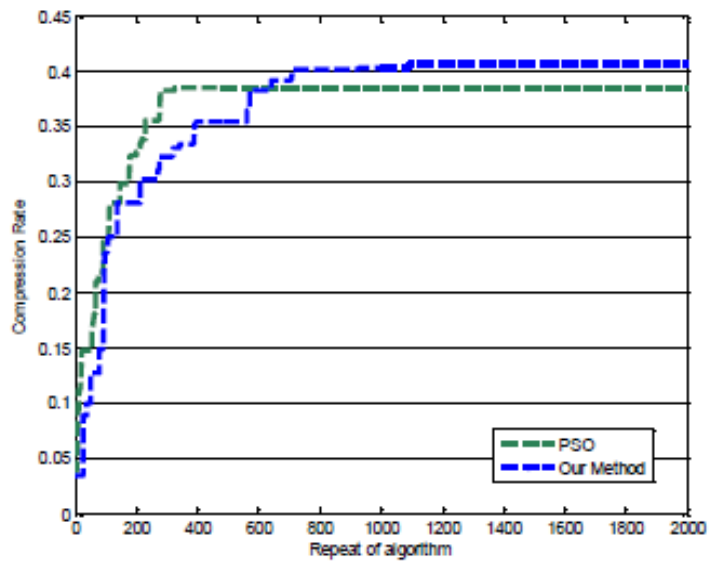

**Figure 16:** Results obtained from the compression rate upon the image F16
(with 16 colors)

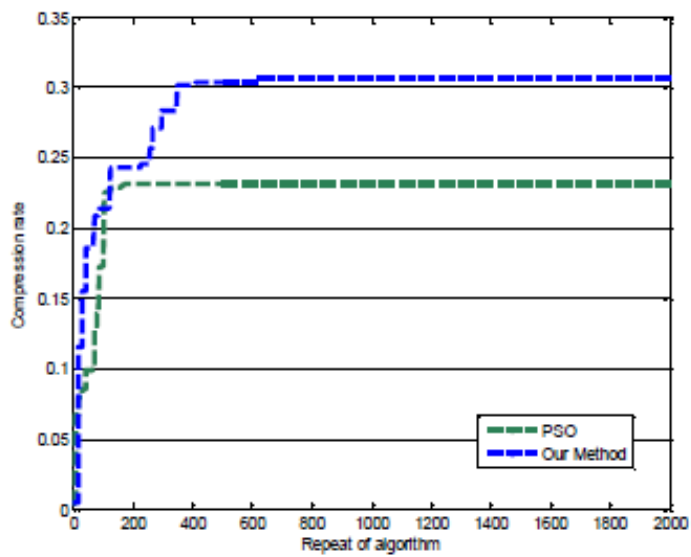

**Figure 17:** Results obtained from the compression rate upon the image Baboon
(with 16 colors)

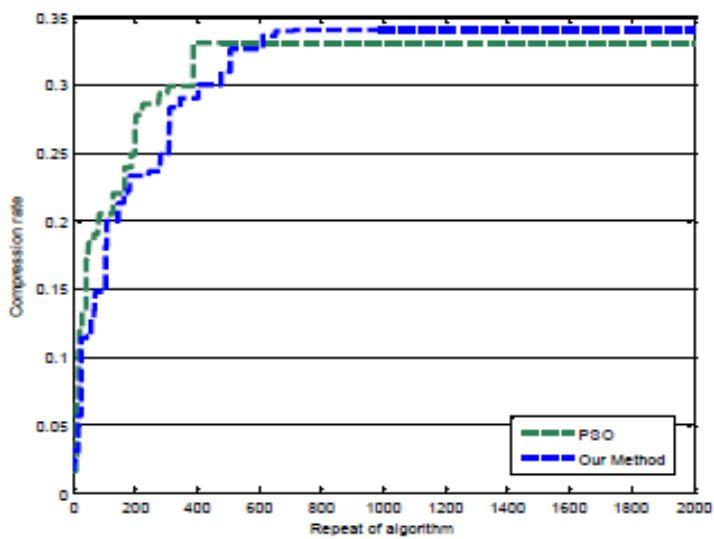

**Figure 18:** Results obtained from the compression rate upon the image Snow
(with 16 colors)

## 5-2 results evaluation according to image entropy

In this part, the results are evaluated according to the entropy of the intended image. The results of this part are also exerted on 64 and 16 colors. Table 1 shows the image entropy when we have 64 colors. Results shown in this table indicate the better efficiency of the suggested algorithm compared to the PSO method. Considering these results, it is observed that the suggested algorithm has a better entropy in all 6 images and this shows that the suggested algorithm has a better efficiency. In table 2, shows the images entropy when the image has 16 different images. Also in this condition, the suggested algorithm can decrease entropy, and it has a better entropy compared to PSO method.

**Table 1:** Entropy obtained from the suggested method and the PSO method
(with 64 colors)

| Images | Suggested method | PSO |
|---|---|---|
| Lena | 171 | 255 |
| Pepper | 173 | 293 |
| Sailboat | 188 | 250 |
| F16 | 186 | 231 |
| Baboon | 187 | 336 |
| Snow | 200 | 345 |

**Table 2:** Entropy obtained from the suggested method and the PSO method
(with 16 colors)

| Images | Suggested method | PSO |
|--------|------------------|------|
| Lena | 356 | 1083 |
| Pepper | 284 | 1034 |
| Sailboat | 311 | 866 |
| F16 | 262 | 986 |
| Baboon | 305 | 952 |
| Snow | 318 | 1041 |

**Conclusion**

In this article, an algorithm based on the imperialist competitive algorithm were introduced for lossless compression of color images. Using the imperialist competitive algorithm, the suggested algorithm tries to optimize the indicators of color map, so that the image compression rate is increased. According to the obtained results which was exerted on 6 images, the suggested method functions better than other methods and can decrease the image size up to 30 to 45 percent.